\documentclass[acmsmall,screen]{acmart}
\usepackage{multirow}
\usepackage{makecell}
\usepackage{subfigure}
\usepackage{mathtools}
\usepackage{amsmath} 
\usepackage{graphicx} 
\usepackage{tikz}
\usepackage{circledsteps}
\usepackage{booktabs}
\usepackage{diagbox}
\usepackage[most]{tcolorbox}
\usepackage[table]{xcolor}
\usepackage{pifont}
\usepackage{tabularx}
\AtBeginDocument{%
  }

\setcopyright{acmlicensed}
\copyrightyear{2018}
\acmYear{2018}
\acmDOI{XXXXXXX.XXXXXXX}
\acmConference[Conference acronym 'XX]{Make sure to enter the correct
  conference title from your rights confirmation email}{June 03--05,
  2018}{Woodstock, NY}
\acmISBN{978-1-4503-XXXX-X/2018/06}

\begin{document}
\newcommand\Aa{{\mathcal{A} }}
\newcommand\Bb{{\mathcal{B} }}
\newcommand\Cc{{\mathcal{C} }}
\newcommand\Ee{{\mathcal{E} }}
\newcommand\Ff{{\mathcal{F} }}
\newcommand\Gg{{\mathcal{G} }}
\newcommand\Jj{{\mathcal{J}}}
\newcommand\Ll{{\mathcal{L}}}
\newcommand\Mm{{\mathcal{M} }}
\newcommand\Nn{{\mathbb{N} }}
\newcommand\Pp{{\mathcal{P} }}
\newcommand\Qq{{\mathcal{Q} }}
\newcommand\Dd{{\mathcal{D} }}
\newcommand\Ss{{\mathcal{S} }}
\newcommand\Tt{{\mathcal{T} }}
\newcommand\transet{{\mathscr{T} }}
\newcommand\Rr{{\mathcal{R} }}
\newcommand\Vv{{\mathcal{V}}}
\newcommand\Zz{{\mathcal{Z} }}

\newcommand\tool{{\textsf{FuncDroid}}}
\newcommand\method{{\textsf{IFO}}}

\newcommand\FDG{{\textsf{FFG}}}
\newcommand\FFDG{{\textsf{FDG}$_\textsf{Func}$}}
\newcommand\WFDG{{\textsf{FDG}$_\textsf{Widg}$}}
\newcommand\Fun{{\textsf{Fun}}}
\newcommand\Data{{\textsf{Data}}}
\newcommand\Path{{\textsf{Path}}}
\newcommand\SFG{{\textsf{SFG}}}
\newcommand\Dep{{\textsf{Dep}}}

\newcommand\control{{\textsf{CntDep}}}
\newcommand\data{{\textsf{DtaDep}}}
\newcommand\inclusion{{\textsf{NclDep}}}
\newcommand\domain{{\textsf{DmnDep}}}

\newtcolorbox{bugbox}[1]{
  enhanced,
  breakable,
  fontupper=\footnotesize,
  colback=red!2,
  colframe=red!45!black,
  colbacktitle=red!12,
  coltitle=black,
  boxrule=0.6pt,
  arc=1.2mm,
  left=1.2mm,
  right=1.2mm,
  top=0.5mm,
  bottom=0.5mm,
  title=\textbf{#1}
}

\newtcbox{\funcbox}{
  on line,
  tcbox raise base,
  boxsep=0pt,
  colback=blue!7,
  colframe=blue!45!black,
  boxrule=0.5pt,
  arc=1.0mm,
  left=0.6mm,
  right=0.6mm,
  top=0.2mm,
  bottom=0.2mm
}

\newtcolorbox{examplebox}{
  enhanced,
  boxrule=0.5pt,
  colframe=black!35,
  colback=black!3,
  arc=2mm,
  left=1.5mm,
  right=1.5mm,
  top=0.5mm,
  bottom=0.5mm
}

\newcommand{\cmark}{\ding{51}}
\newcommand{\xmark}{\ding{55}}

\newcommand{\jinlong}[1]{\color{red} {JL: #1 :LJ}\color{black}}
\newcommand{\yjw}[1]{\color{blue} { Y: #1 :Y }\color{black}}

\title{{\tool}: Towards Inter-Functional Flows for Comprehensive Mobile App GUI Testing}


\author{Jinlong He}

\author{Changwei Xia}

\author{Binru Huang}

\author{Jiwei Yan}

\author{Jun Yan}

\author{Jian Zhang}

\renewcommand{\shortauthors}{Trovato et al.}

\begin{abstract}
As mobile application (app) functionalities grow increasingly complex and their iterations accelerate, ensuring high reliability presents significant challenges.
While functionality-oriented GUI testing has attracted growing research attention, existing approaches largely overlook interactions across functionalities, making them ineffective at uncovering deep bugs hidden in inter-functional behaviors.
To fill this gap, we first design a \textbf{F}unctional \textbf{F}low \textbf{G}raph (\textbf{{\FDG}}), a behavioral model that explicitly captures an app’s functional units and their inter-functional interactions.
Based on the {\FDG}, we further introduce an \textit{inter-functional-flow-oriented GUI testing} approach with the dual goals of precise model construction and deep bug detection.
This approach is realized through a \textit{long–short-term-view-guided testing} process. By combining two complementary test-generation views, it can adaptively refine functional boundaries and systematically explore inter-functional flows under diverse triggering conditions.
We implement our approach in a tool called \textbf{{\tool}}, and evaluate it on two benchmarks: (1) a widely‑used open‑source benchmark with 50 reproducible crash bugs and (2) a diverse set of 52 popular commercial apps. Experimental results demonstrate that {\tool} significantly outperforms state‑of‑the‑art baselines in both coverage (+28\%) and bug detection number (+107\%). Moreover, {\tool} successfully uncovers 18 previously unknown non‑crash functional bugs in commercial apps, confirming its practical effectiveness.
\end{abstract}

\ccsdesc[500]{Software and its engineering~Software testing and debugging}
\keywords{Mobile App, GUI Testing, Functional Bug, Large Language Model}

\received{20 February 2007}
\received[revised]{12 March 2009}
\received[accepted]{5 June 2009}

\maketitle

\section{Introduction}


Mobile applications (apps) are now deeply embedded in modern society, permeating nearly every aspect of daily life.
These apps can be viewed as collections of functionalities, with user interaction primarily consisting of operating these functional components.
As app functionality design grows increasingly complex and functional iterations accelerate, ensuring high reliability poses significant challenges. 
While a wide range of \textit{coverage-oriented} automated GUI testing techniques have been proposed~\cite{monkey, droidbot, time-machine, APE, TrimDroid, Stoat, ComboDroid}, mobile app testing in practice still relies heavily on manual effort.
Testers frequently design functionality-specific test cases to detect defects, highlighting the continued effectiveness of \textit{functionality-oriented testing} in bug discovery.
However, manually designing such functional test cases is time-consuming, labor-intensive, and difficult to scale.
Consequently, a key research challenge in recent years has been to enhance automated GUI testing so that its bug-detection effectiveness can approach that of manual, functionality-oriented testing, while retaining the scalability benefits of automation.

To gain a deeper understanding of mobile apps and enhance traditional GUI testing methods, a growing body of recent research ~\cite{gptdroid,visiondroid,AppAgent,droidbot-gpt,VisionTasker,autodroid} has incorporated large language models (LLMs)~\cite{openai_gpt4o,anthropic_claude,deepseek,qwen,meta_llama}. These studies leverage LLMs' strengths in semantic comprehension and contextual reasoning to enable a degree of \textit{functionality-oriented} test generation, which can be broadly grouped into two categories.
The first category identifies specific functional patterns. Approaches like GPTDroid~\cite{gptdroid} and VisionDroid~\cite{visiondroid} fall into this category. They employ LLMs to directly interpret GUI pages, summarize page sequences into reusable functional patterns, and subsequently apply these patterns to guide further exploration.
The second category focuses on augmenting general function-related app testing knowledge. It comprises frameworks such as LLMDroid~\cite{llmdroid} and MemoDroid~\cite{memodroid}, which enhance existing testing tools (either traditional or LLM-based) by constructing knowledge bases that capture app functionality characteristics. These frameworks distill common interaction patterns and effective exploration strategies from historical testing sessions to guide more efficient exploration.

While these approaches mark a significant shift from \textit{coverage-oriented} to \textit{functionality-oriented} testing, their modeling of functionality is often incomplete and oversimplified, typically represented as a mere linear sequence of events.
More critically, existing approaches primarily operate by testing individual functionalities in isolation, lacking attention to interactions among functionalities. 
Therefore, such approaches often fall short in uncovering deeper bugs that are hidden in the interactions between different functionalities.

To achieve genuinely effective functionality-oriented GUI testing for deep bug detection, we should address the following challenges.
\textbf{First, how can we build an accurate behavioral model that precisely delineates functional boundaries and captures their interactions?}
In practice, functional boundaries are often ambiguous, as functionality can be defined at multiple levels of granularity, ranging from the entire application to individual widget operations.
This ambiguity complicates the automatic segmentation of the GUI into coherent \textit{meaningful functionalities}, since their boundaries may evolve as more knowledge about the app is acquired.
Moreover, as functional boundaries shift, accurately identifying and modeling the flows between functionalities is also challenging.
\textbf{Second, based on such a behavioral model, how can we achieve effective testing that systematically exercises complex interaction paths and increases the likelihood of uncovering deep bugs?}
In real-world apps, the number of meaningful functionalities is often large, and the combinations of functional flows among them lead to an exponential growth of the test space. Without proper test guidance, testing efforts tend to degenerate into redundant and shallow exploration, effectively becoming a blind search that is prone to missing subtle, context-dependent defects.
\textbf{Last but not least}, given that a complete inter-functional model emerges only through iterative exploration, separating model construction from testing is neither practical nor cost-effective.
Thus, it is essential to address \textbf{how to guide the exploration process such that model refinement and testing proceed in a mutually reinforcing manner}, progressively improving the identification of functional boundaries and flows. This issue is a holistic challenge that is tightly intertwined with the two challenges above.

To address the overall challenges, we first design a behavioral model called \textbf{Functional Flow Graph ({\FDG})}, whose node represents a \textit{meaningful functionality}, and edge  represents an \textit{inter-functional flow}. 
Upon the {\FDG}, we further propose an \textbf{inter‑functional-flow-oriented GUI testing approach}, which employs an iterative process that systematically constructs and exploits the {\FDG} to detect deep interaction bugs. 
This approach is mainly realized through a \textit{long–short-term-view-guided testing} process. By combining two complementary test-generation views, it can adaptively refine functional boundaries and systematically explore inter-functional flows under diverse triggering conditions.

Specifically, we perform two synergistic steps in each iteration: the \textit{long-short-term view exploration} and the \textit{functional flow graph updating}.
\textbf{(1) Long-Short-Term View Exploration.} This method employs two complementary perspectives---long-term and short-term---for test scenario generation, each targeting one of the core challenges identified earlier.
To address the first challenge, the \textit{Long‑Term View} generates test scenarios aimed at refining the {\FDG}. These scenarios challenge the completeness and independence of hypothesized meaningful functionalities and examine the precision of flow conditions, thereby producing evidence to enhance the {\FDG} in the next step. 
To address the second challenge, the \textit{Short‑Term View} generates test scenarios for intensive, in-depth exploration based on the current {\FDG}. By applying both \textit{single-flow} and \textit{cross-flow} metamorphic transformations to the state conditions of existing flows, it systematically explores the relevant state subspaces to uncover bugs that arise from complex functional interactions.
\textbf{(2) Functional Flow Graph Updating.} Based on test scenarios generated from both views, corresponding GUI traces are constructed and executed as test cases. The execution results, in turn, support the updating of the functional flow graph. By analyzing observed behaviors and flow conditions, we define three types of \textit{node-updating} operations and four types of\textit{ flow-updating} operations. These updates work in concert with the long–short-term view testing to continuously refine the model and enable systematic, in-depth GUI exploration.

To evaluate the effectiveness of our approach, we implement a tool named \textbf{{\tool}}.
We first conduct extensive experiments on the widely used open-source benchmark \textit{Themis}~\cite{Themis},
which comprises \textbf{50} usable app versions along with their corresponding crash bugs.
On this benchmark, {\tool} significantly outperforms five state-of-the-art baselines in both coverage and bug detection capability.
Specifically, {\tool} achieves a \textbf{28\% improvement in activity coverage} and detects a total of \textbf{62 bugs},
including 49 crash bugs and 13 non-crash functional bugs.
In comparison, the best-performing baseline detects 30 bugs,
indicating that our approach achieves at least \textbf{107\% increase in bug detection effectiveness}.
To further assess its practicality on real-world apps,
we evaluate {\tool} on \textbf{52} popular, high-download commercial apps collected from Google Play and HUAWEI AppGallery.
On their latest versions, {\tool} discovers \textbf{18 previously unknown non-crash functional bugs},
showing its effectiveness in complex, real-world scenarios.


In conclusion, the major contributions of this paper are as follows:
\setlength{\leftmargini}{15pt}
\begin{itemize}
    \item We present the first systematic, functionality-oriented model for mobile apps, called the \textbf{Functional Flow Graph (\FDG)}, which captures both individual functionalities and their inter-functional flows.
    
    \item We propose an \textbf{inter-functional-flow-oriented GUI testing} approach that iteratively constructs and exploits the {\FDG} to systematically explore state subspaces and effectively detect deep bugs arising from complex functional interactions.
    
    \item We implement our approach in a tool {\tool} and evaluate it on both open-source and commercial apps, uncovering \textbf{62 existing bugs} and \textbf{18 previously unknown functional bugs}. The results demonstrate that a precise {\FDG} substantially improves deep bug detection.
\end{itemize}

\section{Motivating Example}\label{sec:motivation}

\begin{figure} [htpb]
\setlength{\abovecaptionskip}{5pt}
\setlength{\belowcaptionskip}{-5pt}
    \centering
     \includegraphics[width=0.98\linewidth]{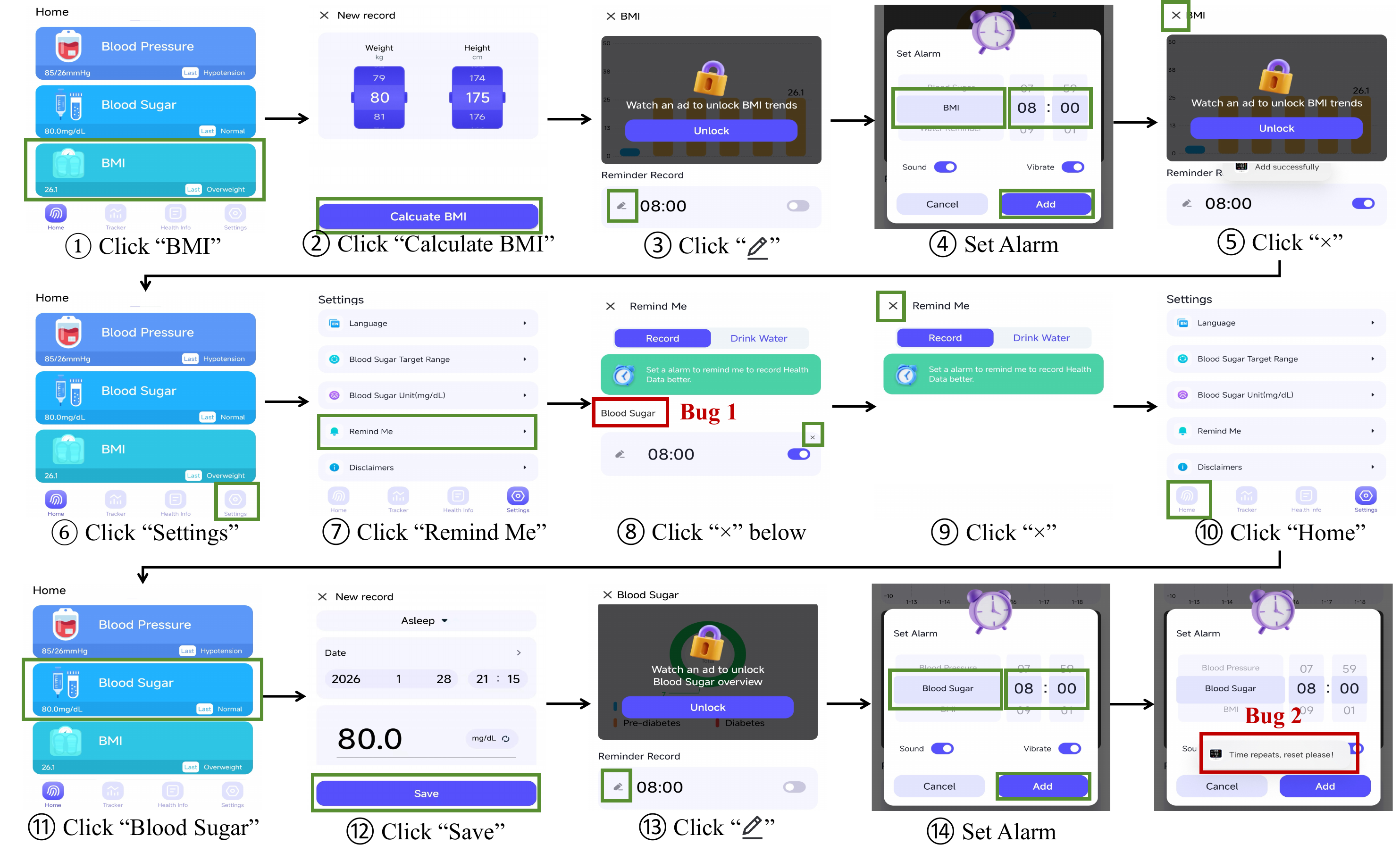}
    \caption{Two Functional Bugs in the Motivating App \textit{Blood Pressure}. 
    }
    \label{fig:bug-example}
\end{figure}

Before presenting our approach, we introduce a motivating example that underscores the need for considering inter-functional flows and exposes the limitations of existing techniques for testing complex cross-functional interactions. Figure~\ref{fig:bug-example} shows two bugs in the \textit{Blood Pressure} app on Google Play, which has been downloaded over 10 million times.
We give the details about the two bugs in the following. Both bugs are triggered only by a specific sequence of actions spanning several distinct functionalities. Testing a single functionality in isolation may therefore miss these bugs. The second bug is particularly subtle, as it requires a precise condition: the same alarm configuration must be deleted in one functional context and then re-added in another, revealing an inconsistency in how the app manages alarm identities across modules. Notably, neither of these two bugs was detected by any of the baseline testing tools.

\begin{bugbox}{Functional Bug~1: BMI Alarm is Misclassified}

\textbf{Tigger Scenario.} Sets a new BMI alarm in the app.

\textbf{GUI Trace (Steps).}
\begin{enumerate}
    \item On the ``Home'' page, click ``BMI'' to set the BMI value (\Circled{1}), click ``Calculate BMI'' to save it (\Circled{2}), then click ``Edit'' to set an alarm (\Circled{3}).
    \item Choose ``BMI'' and ``08:00'', then click ``Add'' (\Circled{4}).
    \item Click the page close icon ``$\times$'' to go back to the ``Home'' page (\Circled{5}), tap ``Settings'' (\Circled{6}), and click ``Remind Me'' to view alarms.
\end{enumerate}

\textbf{Observed Bug.} The ``Remind Me'' page shows an alarm for ``blood sugar'' instead of ``BMI''.
\end{bugbox}

\begin{bugbox}{Functional Bug~2: Re-adding a Deleted Alarm Fails}
\textbf{Trigger Scenario.} Re-adds a Blood Sugar alarm at the same time as a previously deleted one.

\textbf{GUI Trace (Steps).}
\begin{enumerate}
    \item On the ``Remind Me'' page, click the alarm delete button``$\times$'' (\Circled{8}), then click the page close icon ``$\times$'' (\Circled{9}) to return to the ``Settings'' page, and tap ``Home'' (\Circled{10}) to go back to the ``Home'' page.
    \item Click ``Blood Sugar'' on the ``Home'' page (\Circled{11}), click ``Save'' (\Circled{12}), then click ``Edit'' (\Circled{13}) to set an alarm.
    \item Choose ``Blood Sugar'' and ``08:00'', then click ``Add'' (\Circled{14}).
\end{enumerate}

\textbf{Observed Bug.} The app shows a ``Time repeat'' warning message, and the alarm is not added.
\end{bugbox}


\section{Approach of {\tool}}


To uncover complex bugs such as those illustrated in Figure~\ref{fig:bug-example}, we design the \textbf{Functional Flow Graph ({\FDG})} to model functional interactions
in terms of the state conditions under which they occur, thereby enabling in-depth cross-functionality testing.
In the {\FDG}, each node represents a meaningful functionality, and each edge, denoted as $n \xrightarrow{(\pi, \phi, \pi')} n'$, represents a flow: it specifies that after executing the trace $\pi$ of functionality $n$, if the app state satisfies the condition $\phi$, then the trace $\pi'$ of functionality $n'$ becomes executable. Crucially, each such flow, through its condition $\phi$, defines a meaningful subset of the overall app state space. By constructing a graph of these flows, the {\FDG} thus provides a structured partition of the state space into coherent, functionally relevant regions. {\tool}'s key idea is that, \textit{guided by this partition, testing can then focus on thorough exploration within each state subset, rather than attempting a blind, exhaustive search of the entire space.} Based on this idea, we propose an \textbf{Inter‑Functional-Flow-oriented GUI testing approach}, which iteratively builds and exploits the {\FDG} to cover state subspaces systematically and detect deep interaction bugs effectively.

\subsection{Definition of Functional Flow Graph}\label{fdg:define}

While the concept of functionality is frequently mentioned in GUI testing research, its practical definition remains ambiguous. This ambiguity arises because functionality can be conceptualized at multiple granularity levels, spanning from fine-grained widget-level to coarse-grained app-level, with no universally accepted standard.
In this work, we argue that a \textit{meaningful functionality} is the most pertinent level of granularity for testing. A meaningful functionality, as defined here, must satisfy two key criteria: 
(1) it is \textbf{complete}, i.e., it accomplishes a well-defined, user-centric task; 
and (2) it is \textbf{independent}, i.e., it is logically indivisible, meaning it does not itself contain other meaningful sub-functionalities.

Based on the concept of ``meaningful functionality'', we accordingly define the \textbf{functional flow graph ({\FDG})} as a directed graph $ G = (N, E) $, where $ N $ denotes the set of meaningful functionalities and $ E $ represents the inter-functional flows (\textit{flow} for short) between them. Specifically:

\begin{itemize} 
\item A \textit{functionality} $n\in N$ is defined as a triple $(\textit{goal}, \textit{vars}, \textit{traces})$, 
where:
\begin{itemize}
\item $\textit{goal}$ is the semantic intent of the functionality (e.g., “Alarm Management”);
\item $\textit{vars}$ is the set of state variables that are accessed or modified during the execution of the current functionality, spanning GUI property, data entity, and system environment setting dimensions; 
\item $\textit{traces}$ is a set of execution traces, each consisting of a sequence of widget actions involved in the functionality.
\end{itemize}
\item An \textit{flow}  $ (n, n') \in E $ is annotated with a triple $(\pi, \phi, \pi')$. This denotes that executing the trace $\pi$ of functionality $n$ leads the app state to satisfy the condition $\phi$, which enables the activation of the trace $\pi'$ in functionality $n'$.
Formally, we write it as $n \xrightarrow{(\pi,\phi,\pi')} n'$.
\end{itemize}

\subsection{Inter-Functional-Flow-oriented GUI Testing}
Based on that model design, we then propose an inter‑functional-flow-oriented GUI testing approach, which targets covering more flows to detect deep bugs. As shown in Figure~\ref{overview}, it first constructs the initial functional flow graph ({\FDG}) from the execution trace collected by the lightweight automated GUI exploration (Section~\ref{sec:init}). 
Then it employs an iterative process to systematically uncover deep interaction bugs by exploring the state subspaces defined by the {\FDG}. Each iteration performs two synergistic steps:

\begin{figure}[!htbp]
\setlength{\abovecaptionskip}{5pt}
\setlength{\belowcaptionskip}{-5pt}
    \centering
    \includegraphics[width=1\textwidth]{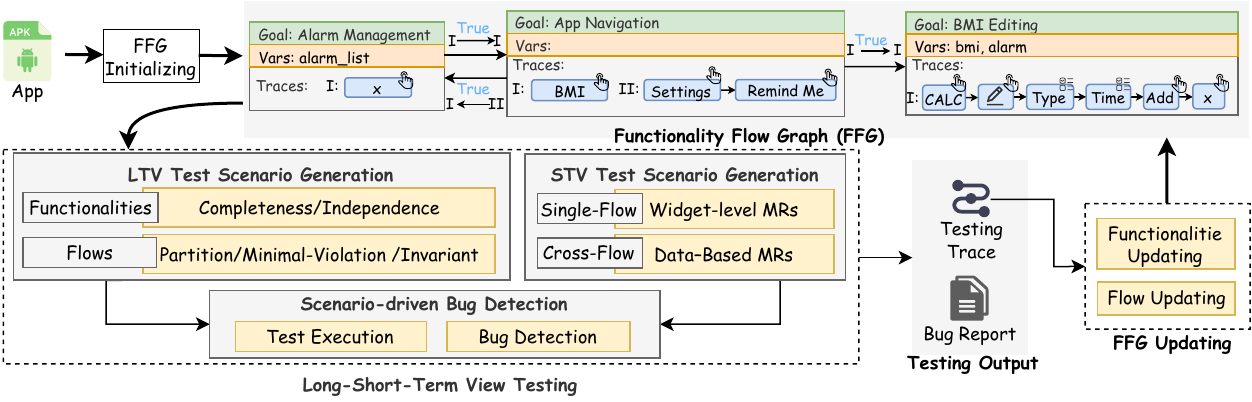}
    \caption{Overview of {\tool}}
    \label{overview}
\end{figure}

\begin{enumerate}
\item \textit{Long‑Short-Term View Exploration (Section~\ref{sec:lstv})}. It is composed of three coordinated phases: 
(a) \textit{Long‑Term View (LTV) test Scenario Generation} generates tests that target refining the {\FDG} model; 
(b) \textit{Short‑Term View (STV) Test Scenario Generation} generates tests aimed at intensive exploration of current flow hypotheses; and 
(c) \textit{Scenario‑driven Bug Detection} performs both test execution and bug detection; 
    
\item \textit{Functional Flow Graph Updating (Section~\ref{sec:update})}. It continuously carries out the actual functionality and flow updates on the {\FDG} based on the collected execution information. 
\end{enumerate}

Through the iterative cycle with both steps, {\tool} progressively deepens its understanding of the app while efficiently covering diverse and effective flows.


\subsubsection{Functional Flow Graph Initializing}\label{sec:init}
This step aims to initialize the {\FDG}. It begins with a lightweight, automated GUI exploration (e.g., using random or systematic exploration strategies) to collect an initial execution trace, which serves as the basis for hypothesizing an initial set of functionalities and flows among them. 

\begin{figure}[b!]
    \centering
    \subfigure[The Initial Execution Trace.]{
        \label{fig:ffg-init-trace}
        \centering
        \includegraphics[width=0.73\linewidth]{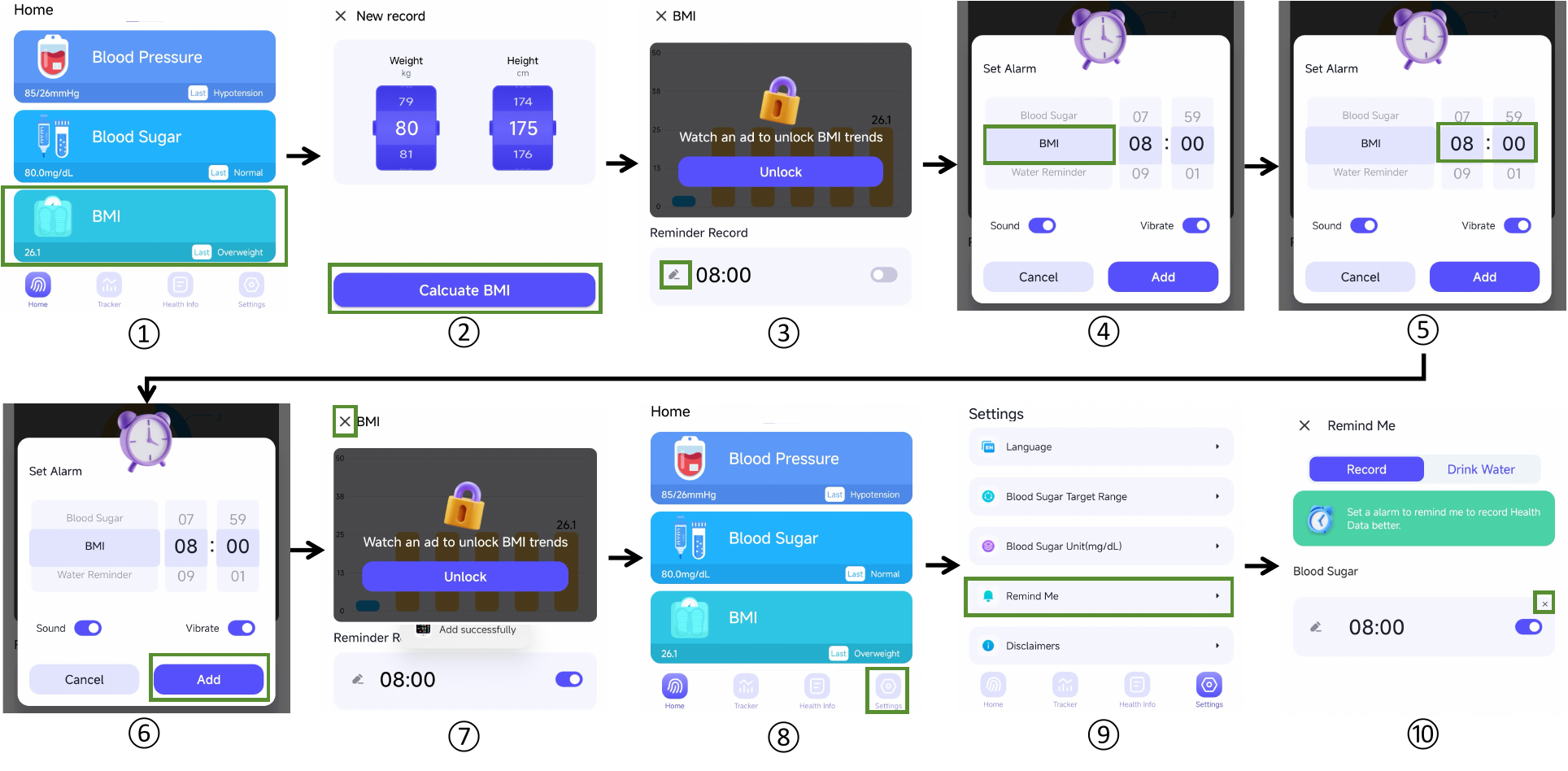}
        }
    \begin{tikzpicture}
        \draw[dashed] (0,-3cm) -- (0, 2cm); 
    \end{tikzpicture}
    \subfigure[The Initial {\FDG}.]{
        \label{fig:ffg-init-graph}
        \centering
        \includegraphics[width=0.20\linewidth]{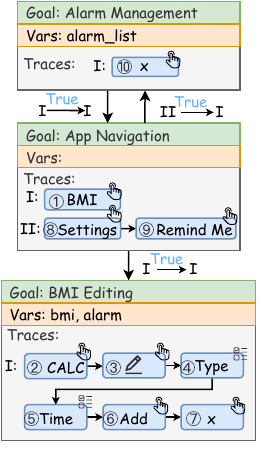}
        }
    \vspace{-0.3cm}
    \caption{{\FDG} Initialization of the \textit{Blood Pressure} App.}    \label{fig:ffg-init}
\end{figure}
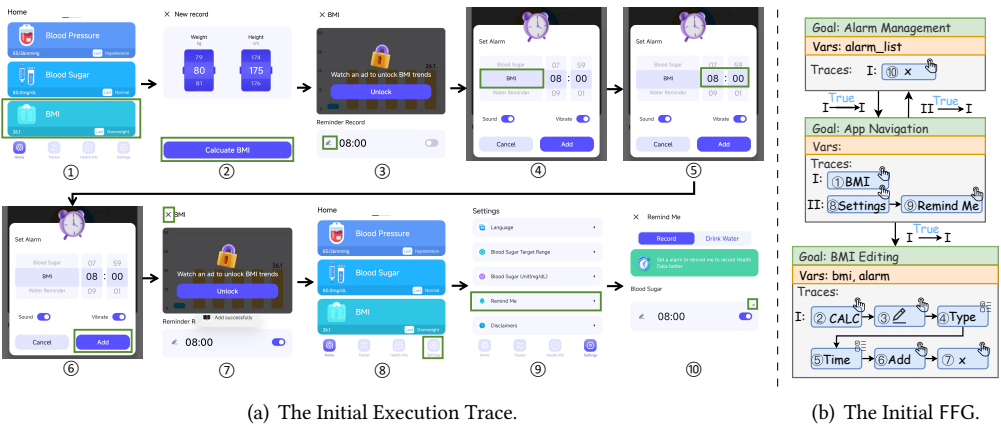

First, given the execution trace, we partition it into a set of candidate meaningful functionalities based on semantic coherence. 
For each GUI page visited in the trace, we extract its screenshot, widget hierarchy tree, and textual content, and analyze them using a multimodal large language model (MLLM)
to infer the page’s \textit{goal}. Meanwhile, we identify the state variables \textit{vars} (e.g., GUI properties, data entities, or system environment settings) that are accessed or modified on that page.
Specifically, starting from the first page in the trace, we initialize a candidate functionality $n = (goal, vars, traces)$, where \textit{traces} initially contains only that page. 
As the trace proceeds in order, we compute the goal of each subsequent page and encode it as a semantic embedding, which is then compared with the goal of the currently candidate functionality.
If the semantic similarity exceeds a predefined threshold, the page is regarded as a continuation of the current functionality: its actions are appended to \textit{traces}, and newly identified variables are merged into \textit{vars}.
Otherwise, if the page exhibits a significantly different goal, we can initialize a new candidate functionality with that page as its starting point.

Then, after the functionalities are partitioned, we extract inter-functional flows at functionality switching points and add them to the initial {\FDG}.
Specifically, given the initial execution trace, if a trace $\pi$ belonging to a functionality $n$ is immediately followed by a trace $\pi'$ belonging to another functionality $n'$, we create a candidate edge $n \xrightarrow{(\pi, \phi, \pi')} n'$ 
At this stage, the state condition $\phi$ is conservatively set to $\textit{True}$, indicating that the transition is inferred solely from the observed temporal succession, without yet reasoning about the precise state conditions that enable it.

For example, Figure~\ref{fig:ffg-init} illustrates the initialization process on the \textit{Blood Pressure} app.
Its initial execution trace (Figure~\ref{fig:ffg-init-trace}) consists of ten actions \Circled{1}-\Circled{10}. 
\begin{tcolorbox}[
  enhanced,
  fontupper=\footnotesize,
  breakable,
  colback=gray!3,
  colframe=black!35,
  colbacktitle=black!8,
  coltitle=black,
  boxrule=0.6pt,
  arc=1.2mm,
  left=1.2mm,
  right=1.2mm,
  top=0.8mm,
  bottom=0.8mm,
  title=\textbf{The Initial Execution Trace with 10 Actions}
]
\Circled{1} clicking the ``BMI'' button on the ``Home'' page;
\Circled{2} clicking the ``Calculate BMI'' button on the reached ``BMI'' page;
\Circled{3} clicking the ``Edit'' button on the ``BMI'' page;
\Circled{4} configuring alarm type on the reached ``Alarm Editing'' page;
\Circled{5} configuring alarm time on the ``Alarm Editing'' page;
\Circled{6} clicking the ``Add'' button on the ``Alarm Editing'' page;
\Circled{7} clicking the close button ``$\times$'' on the ``BMI'' page;
\Circled{8} clicking the ``Settings'' button on the ``Home'' page;
\Circled{9} clicking the ``Remind me'' button on the reached ``Settings'' page;
\Circled{10} clicking the ``$\times$'' button on the reached ``Remind me'' page.
\end{tcolorbox}

By analyzing the \textit{goal} of each page, the initial {\FDG} is constructed as shown in Figure~\ref{fig:ffg-init-graph}. The trace is partitioned into three meaningful functionalities. Specifically, the functionality \funcbox{\textit{App Navigation}} encompasses two separate traces: trace~\textup{I} containing action \Circled{1}, and trace~\textup{II} containing actions \Circled{8} and \Circled{9}; the functionality \funcbox{\textit{BMI Editing}} encompasses one trace~\textup{I} containing actions \Circled{2}--\Circled{7}; the functionality \funcbox{\textit{Alarm Management}} encompasses one trace~\textup{I} containing action \Circled{10}. The conditions of the flows between these three functionalities are initially set to be \textit{True}.



\subsubsection{Long-Short-Term View (LSTV) Exploration}\label{sec:lstv}
This step implements our core testing strategy by systematically generating inter-functional test scenarios from two complementary perspectives: the \textit{Long-Term View} (LTV) and the \textit{Short-Term View} (STV) .
Although these two views differ in design intent, the scenarios they produce are uniformly consumed by a common \textit{Scenario-driven Bug Detection} phase.
To support this unified processing, we introduce a formal representation for scenarios generated from both views, referred to as \textit{LSTV test scenarios}.
An \textit{LSTV test scenario} is defined as a 4-tuple $(\textit{type}, \textit{strategy}, \textit{object}, \textit{guidance})$, where 
$\textit{type} \in \{\textsf{LTV}, \textsf{STV}\}$ denotes the perspective of the scenario generation; 
$\textit{strategy}$ indicates the specific generation strategy or metamorphic relation applied;
$\textit{object}\in N\cup E$ specifies the targeted functionality or functional flow to which the strategy is applied;
and $\textit{guidance}$ is a natural‑language description that explains how the $\textit{strategy}$ is instantiated and applied to the $\textit{object}$ in order to construct the scenario.

\paragraph{\textbf{(1) Long-Term View Test Scenario Generation.}}
Since the effectiveness of testing depends on the quality of the underlying model, i.e., the {\FDG}, the long-term view focuses on generating test scenarios that iteratively refine the {\FDG} for subsequent testing.
This phase generates test scenarios that can refine the two core aspects of the {\FDG}: its \textit{functionality definitions} and \textit{flow conditions}.

On one hand, to \textbf{refine functionality definitions}, we design test scenarios that can validate whether each recorded functionality $n \in N$ satisfies the criteria of a \textit{meaningful functionality}, i.e., it is both complete and independent, with the following two strategies.

\begin{figure}[htbp]
\setlength{\abovecaptionskip}{5pt}
\setlength{\belowcaptionskip}{-5pt}
    \centering
     \includegraphics[width=\linewidth]{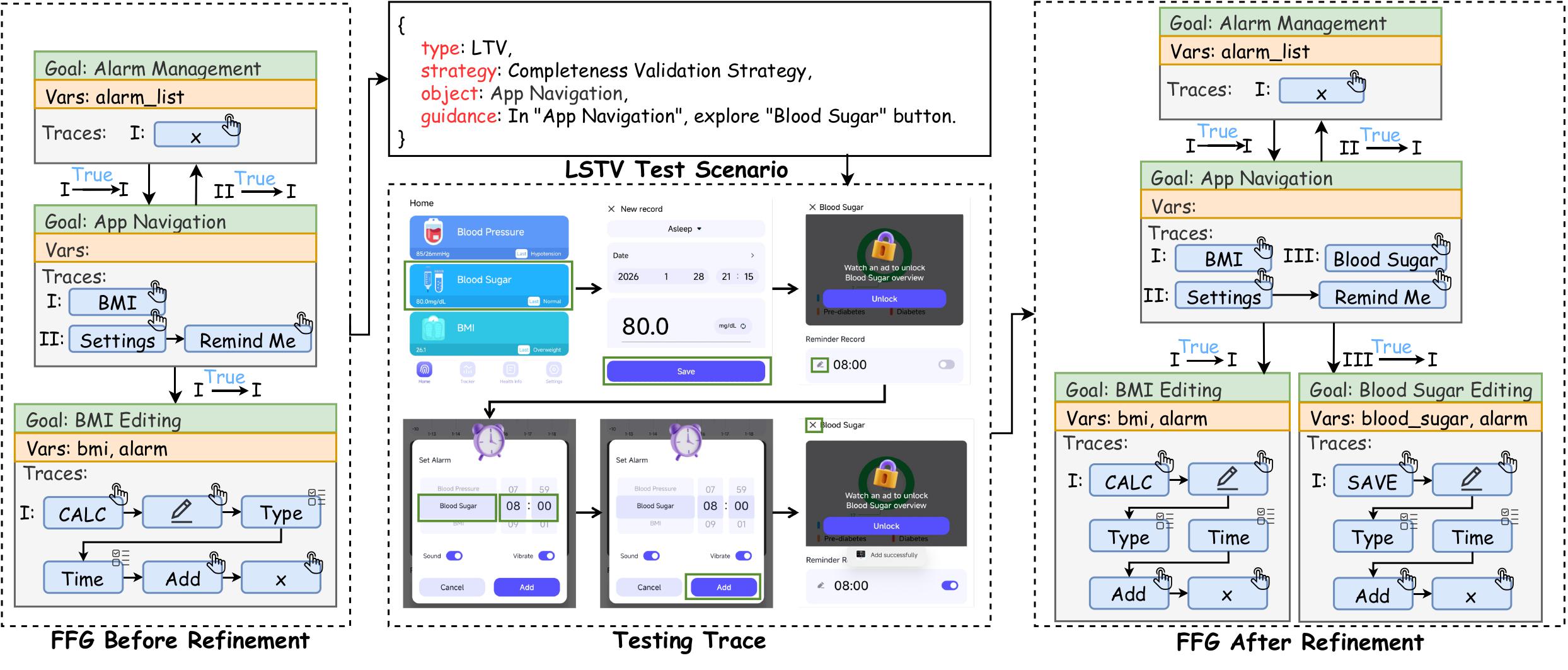}
    \caption{Example of \textsf{LTV} Test Scenario for Refining Functionality Definitions.}
    \label{fig:ifo-update-func}
\end{figure}

\begin{itemize}
    \item \textit{Completeness Validation Strategy}.    
    To challenge the \textit{completeness} of a functionality $n$, we analyze its high-level semantic \textit{goal} together with the GUI pages in its \textit{traces} to identify potential gaps between the intended functionality and the recorded execution. 
    Based on this analysis, we infer a set of potential widget actions that are semantically essential for achieving the \textit{goal} but are absent from the recorded \textit{traces}. 
    We then generate a new test scenario that executes the existing trace of $n$ and subsequently attempts to execute the missing actions.

    
    \item \textit{Independence Validation Strategy}. 
    To challenge the \textit{independence} of a functionality $n$, we compute its \textit{trace-level goals} by generating a semantic goal embedding for each trace in the set \textit{traces}.
    These trace-level goals are then clustered to identify potential subgroups.
    If the clustering reveals a clear separation of traces into groups with low inter-cluster similarity, it indicates that $n$ may encompass multiple semantically coherent sub-tasks.
    For each trace cluster, we generate a new test scenario that executes only the traces within the cluster, in order to validate whether the corresponding ``sub-task'' can be executed as an independent functionality.
\end{itemize}
\begin{examplebox}
For example, Figure~\ref{fig:ifo-update-func} illustrates how a test scenario is generated and used to refine functionality definitions through the completeness validation strategy. We take the {\FDG} shown in Figure~\ref{fig:ffg-init-graph} as the current {\FDG} in this example. We first observe that the action of clicking the ``Blood Sugar'' button (same for ``Blood Pressure'', omitted here for brevity) is not present in the traces of the functionality \funcbox{\textit{App Navigation}}. Consequently, we generate a scenario that attempts to interact with the ``Blood Sugar'' button in order to assess whether the functionality \funcbox{\textit{App Navigation}}  is complete. Using the technique described in the \textit{Scenario‑driven Bug Detection phase}, we can finally get an execution trace that includes the missing actions, which successfully explores the path after clicking that ``Blood Sugar'' button. By goal comparison, this exploration didn't update the existing functionality.
Instead, it leads to the discovery of a new, independent ``Blood Sugar Editing'' functionality.

\end{examplebox}

On the other hand, to \textbf{refine flow conditions}, we design a set of test scenarios that aim to sharpen the flow condition $\phi$ of a flow
$e = n \xrightarrow{(\pi, \phi, \pi')} n'$.
The core intuition is to determine whether the condition $\phi$ can be soundly weakened or strengthened while preserving the transition to $n'$.
Moreover, each condition $\phi$ is expressed in disjunctive normal form (DNF), i.e., a disjunction of conjunctive clauses (e.g., $\phi = \phi_1 \lor \cdots \lor \phi_n$ where $\phi_i = \psi_1 \land \cdots \land \psi_{k_i}$ for each $i$).
To handle the \textit{True} condition, we construct a DNF by analyzing the state variables $\textit{vars}$ of the target functionality $n'$.
To this end, we construct environment configurations that are highly correlated with $\pi'$ through a series of targeted operations.
Executing $\pi'$ under these constructed environments yields a set of new test scenarios.
These strategies are described in detail below.

\begin{figure}[htbp]
\setlength{\abovecaptionskip}{5pt}
\setlength{\belowcaptionskip}{-5pt}
    \centering
     \includegraphics[width=\linewidth]{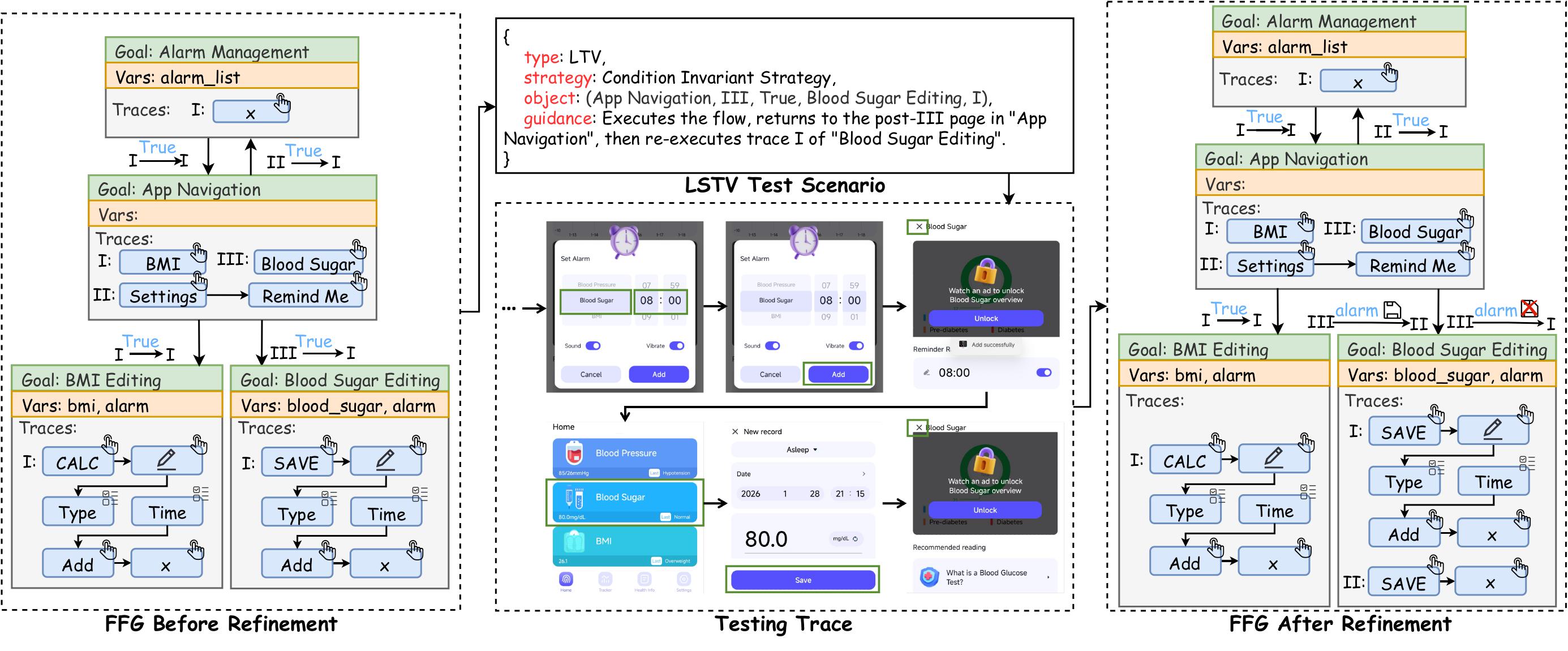}
    \caption{Example of \textsf{LTV} Test Scenario for Refining Flow Conditions.}
    \label{fig:ifo-update-flow}
\end{figure}

\begin{itemize}

\item \textit{Condition Partition Strategy}. 
To assess whether a flow condition $\phi$ should be split into multiple distinct conditions, we generate test scenarios to examine its internal heterogeneity.
Since $\phi$ is expressed as a disjunction (e.g., $\phi_1 \lor \cdots \lor \phi_n$), it can be partitioned into $n$ groups. 
Each resulting sub-condition should be mutually exclusive and semantically meaningful.
For each sub-condition $\phi_i$, we create test scenarios that execute $\pi$ and drive the system into states satisfying $\phi_i$. 
If the outcomes differ across sub-conditions (e.g., $\pi_i$ is activated under $\phi_i$ but leads to $\pi_j$ under $\phi_j$), this indicates that the original flow $e$ should be split into multiple, more precise flows with distinct conditions.


\item \textit{Condition Minimal-Violation Strategy}. 
To assess whether the flow condition $\phi$ can be weakened, we generate test scenarios that execute $\pi$ from states that minimally violate $\phi$. 
To achieve this, we first identify the atomic minimal violation units in $\phi$.
For example, if there exists any clause in the form of $\psi_1 \land \psi_2$ in $\pi$, $\psi_1$ and $\psi_2$ are both taken as minimal violation conditions.
For each minimal violation condition $\psi_i$, a test scenario is created in which $\psi_i$ is false while all other sub-conditions hold. 
If $\pi'$ can still be triggered with a consistent result, it indicates that $\psi_i$ is an unnecessary constraint, suggesting that $\phi$ can be generalized by removing $\psi_i$; conversely, consistent execution reinforces the necessity of $\psi_i$.

\item \textit{Condition Invariant Strategy}. 
To assess whether the flow condition $\phi$ can be strengthened, we generate test scenarios of the form $\pi \cdot \pi' \cdot \alpha \cdot \pi$.
In such a scenario, the core flow $\pi \cdot \pi'$ is executed first.
Next, an intermediate action $\alpha$ is performed to deliberately perturb the resulting app state.
The action $\alpha$ is selected from the traces of $n'$ that can navigate back to a page where $\pi$ is executable, while ensuring that the condition $\phi$ is still satisfied.
Finally, $\pi$ is executed again.
By comparing the outcomes of the two executions of $\pi$, we can determine whether the second execution leads to divergent behavior (e.g., triggering a different $\pi''$ or causing a failure), which would indicate that the current condition $\phi$ is overly permissive or incomplete.
\end{itemize}
\begin{examplebox}
For example, Figure~\ref{fig:ifo-update-flow} illustrates how a test scenario is generated and used to refine flow conditions through the condition invariant strategy. We take the refined {\FDG} shown in Figure~\ref{fig:ifo-update-func} as the current {\FDG} in this example. We then apply the condition invariant strategy to the flow $e = {\sf "App\ Navigation"} \xrightarrow{(\textup{III}, \textrm{True}, \textup{I})} {\sf "Blood\ Sugar\ Editing"}$. This strategy generates a test scenario that first executes the flow, then navigates back to the page reached after trace \textup{III} in \funcbox{\textit{App Navigation}}, and finally attempts to re‑execute trace \textup{I} of \funcbox{\textit{Blood Sugar Editing}}. Following the technique described in the \textit{Scenario‑driven Bug Detection} phase, we obtain the corresponding test trace. The execution reveals that trace \textup{I} cannot be successfully triggered, more specifically, an alarm cannot be added after clicking the ``Save'' button. Since trace \textup{I} of \funcbox{\textit{Blood Sugar Editing}} is designed to add an alarm, and this action cannot be re‑executed, it indicates that the current condition $\textit{True}$ is too weak. Consequently, update the {\FDG} by replacing the condition of flow $e$ with the stronger condition ``$alarm$ is not added''.
\end{examplebox}

Overall, the above three \textsf{LTV} testing strategies systematically generate contrastive test scenarios that compare the flow’s behavior across intentionally varied execution histories, which are prepared for the subsequent \textit{{\FDG} Updating} step.
Nevertheless, the test cases derived from these scenarios can also be used for bug detection.



\paragraph{\textbf{(2) Short-Term View Test Scenario Generation.}}
Different from \textsf{LTV}, this type of view targets directly uncovering deep bugs through exploring the {\FDG}, so it 
focuses on generating test scenarios that intensively probe current flow hypotheses.
This phase employs a metamorphic testing approach, constructing scenarios by applying state transformations (metamorphic relations, MRs) to the variables appearing in a flow’s condition $\phi$. 
The MRs are organized into two complementary levels: those operating within a single flow and those spanning multiple flows based on data dependencies.

First, we design a set of \textbf{single‑flow metamorphic relations}. For a given flow $e = n \xrightarrow{(\pi, \phi, \pi')} n'$, we generate alternative execution paths that still satisfy $\phi$ by altering the trace $\pi$ within functionality $n$. This is achieved by applying widget‑level MRs to the GUI actions in $\pi$. For instance, we may use the following transformations in single‑flow MRs:
\begin{itemize}
\item \textit{Hide/Show}. Hide or show dynamic GUI elements before interacting with them.
\item \textit{Change}. Change the order of filling form fields when order independence is semantically plausible.
\item \textit{Toggle}. Toggle optional GUI components (e.g., \texttt{switches}, \texttt{checkboxes}) that are not strictly required for $\phi$ but may affect the resulting state.
\end{itemize}
Their goal is to generate a variant execution trace $\pi_{\text{var}}$ that, when executed, leads to a state satisfying $\phi$ and thus enables $\pi'$.
In the subsequent phase, if executing this scenario fails to enable $\pi'$, the scenario may be treated as a potentially buggy case.
This probes the robustness of the flow against minor variations in how its condition is established.

Then, we design a set of \textbf{cross‑flow metamorphic relations}. This level targets pairs of functionalities $(n, n')$ that have no direct flow in the current {\FDG} but share related data $d$, i.e., their variable sets $\textit{vars}$ and $\textit{vars}'$ intersect. For each such pair, we identify an existing flow whose target functionality is $n'$ and whose condition is related to the shared data $d$. We then construct scenarios in which a trace of $n$ is executed to transform the data state variables so that they satisfy the condition $\phi$. 
%
Typical transformations in cross‑flow MRs include:
\begin{itemize}
    \item \textit{Create/Delete}. Node $n$ executes a trace to create or delete the data entity $d$.
    \item \textit{Modify‑Attribute}. Node $n$ executes a trace to alter the attributes of the data entity $d$.
    \item \textit{Consume/Produce}. Node $n$ executes a trace to alter the availability of a shared system or data resource associated with the data entity $d$.
\end{itemize}
The resulting scenarios evaluate how indirect data interactions, which lie outside the currently modeled flows, influence the behavior of the target flow $e$.

\begin{figure}[htbp]
\setlength{\abovecaptionskip}{5pt}
\setlength{\belowcaptionskip}{-5pt}
    \centering
     \includegraphics[width=\linewidth]{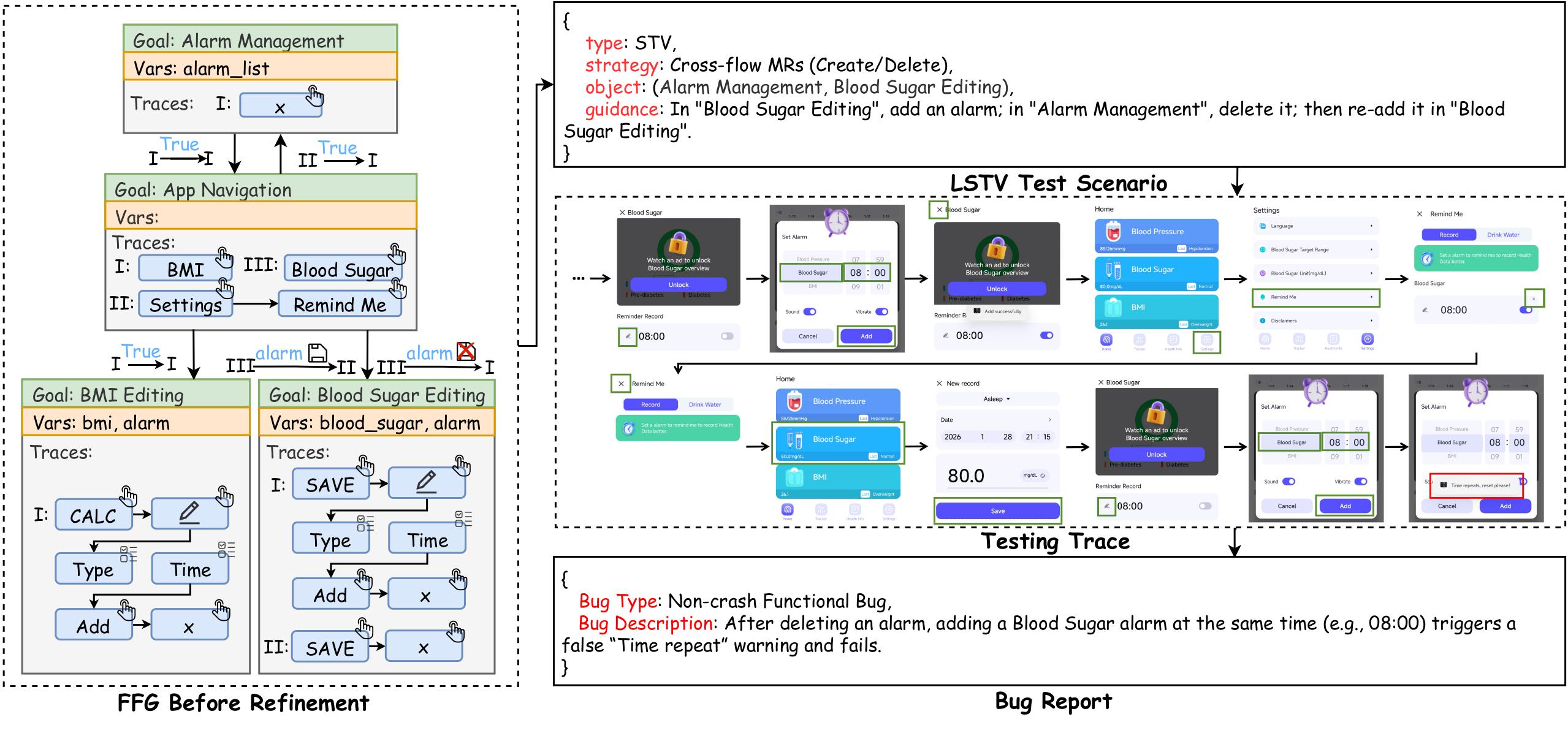}
    \caption{Example of \textsf{STV} Test Scenario for Cross-Flow Metamorphic Relations.}
    \label{fig:ifo-mr}
\end{figure}

\begin{examplebox}
For example, Figure~\ref{fig:ifo-mr} illustrates how a test scenario is generated and used to uncover deep bugs through cross-flow metamorphic transitions. Taking the refined {\FDG} from Figure~\ref{fig:ifo-update-flow} as the current model, we observe that the functionalities \funcbox{\textit{Blood Sugar Editing}} and \funcbox{\textit{Alarm Management}} share data entities (e.g., \textit{alarm} and \textit{alarm\_list}). We therefore locate a flow $e = {\sf "App\ Navigation"} \xrightarrow{(\textup{III}, \phi, \textup{I})} {\sf "Blood\ Sugar\ Editing"}$ where $\phi$ is ``$alarm$ is not added'', and apply the \textit{Create/Delete} MR. The \textit{guidance} of this scenario is: (1) In \funcbox{\textit{Blood Sugar Editing}}, add an alarm; (2) in \funcbox{\textit{Alarm Management}}, delete it; and (3) re-add it in \funcbox{\textit{Blood Sugar Editing}}. By executing tests under this scenario, we successfully detect \textbf{Bug~2} shown in Figure~\ref{fig:bug-example}. 
\end{examplebox}


Overall, the \textsf{STV} scenarios are primarily designed for bug detection; however, executing the test cases derived from them can also contribute to the \textit{{\FDG} Updating} in the next subsection.



\paragraph{\textbf{(3) Scenario‑driven Bug Detection}}

This phase executes the testing traces guided by the high-level \textit{LSTV test scenarios} and then performs bug detection upon them.
First, \textbf{test execution} is performed to execute the testing trace guided by an LSTV test scenario $(\textit{type}, \textit{strategy}, \textit{object}, \textit{guidance})$. Beginning from the main page, the MLLM consults the current {\FDG} to navigate toward the target functionality or flow. It then incrementally translates the high‑level $\textit{guidance}$ into a sequence of widget actions. Each action is then attempted in turn. If an execution fails (e.g., the widget is missing, disabled, or its properties do not match the expected state), the failure is logged, and the MLLM uses the real‑time screenshot along with the {\FDG} to propose a recovery step or an alternative action.
Throughout test execution, \textbf{bug detection} is performed to identify non-crash functional bugs. Unlike crash bugs, which manifest as obvious runtime failures, non‑crash bugs lack a built‑in failure signal and therefore require an explicit oracle to decide whether an observed behavior constitutes a defect. To this end, we adopt the approach established in prior work~\cite{visiondroid}, in which an MLLM is augmented with a knowledge base of common non‑crash bug patterns. Based on this knowledge base, the expected behavior for the testing trace can be inferred. Then we can analyze the final screenshot, widget hierarchy tree, testing trace, and system logs to compare the observed outcome with the expected one.
Finally, it yields a structured bug report, which, together with the execution trace, is fed back to drive model refinement.

\subsubsection{Functional Flow Graph Updating}\label{sec:update}

During the \textit{LSTV Testing} phase, the newly generated test cases are executed.
This process produces additional execution traces, which are collected to iteratively refine our {\FDG} model, progressively making it a more accurate and complete representation of the app.
Within the {\FDG}, two types of targets need to be refined: the set of functionalities $N$ and the set of flows $E$. We discuss each in turn.

First, \textbf{functionality updating} is performed by analyzing the semantic similarity between the \textit{goal} of each newly generated testing trace and those of existing functionalities.
This process is similar to that used in \textit{{\FDG} Initializing}, where semantic similarity is evaluated via embedding comparison. 
However, while \textit{{\FDG}  Initializing} only identifies new functionalities (i.e., the functionality \textit{create} operation) by analyzing traces, \textit{{\FDG} Updating} can further determine whether a new trace triggers a functionality \textit{merge} or \textit{split} operation. Here are all the functionality-related operations used.

\begin{itemize}
    \item \textit{New Functionality Creation}. If a testing trace accomplishes a distinct task whose goal is not similar to any existing functionality, a new functionality node is created and added to $N$.
    
    \item \textit{Existing Functionality Merging}. If a testing trace exhibits a goal and manipulates a set of state variables that align closely with an existing functionality $n$, it is merged into $n$. This expands $n$'s \textit{traces} set and may update its \textit{vars} set, reinforcing the functionality's completeness.
    
    \item \textit{Existing Functionality Splitting}. If a testing trace reveals that the \textit{traces} of an existing functionality $n$ can be partitioned into subsets that accomplish semantically independent, user-centric tasks, $n$ is split into multiple finer-grained functionalities. This refines the model’s granularity to hold the independence criterion.
\end{itemize}

%


Then, \textbf{flow updating} is performed by examining the relationship between a newly generated testing trace and existing flows, with a particular focus on comparing their state conditions.
To identify the state conditions of a new trace, we first locate its functionality-changing points.
After the previous phase, each testing trace is segmented into contiguous trace segments, denoted as $\pi$ and $\pi'$, where $\pi$ is labeled as belonging to functionality $n$, and the subsequent segment $\pi'$ belongs to functionality $n'$.
At each functionality-changing point, we further query the MLLM to identify the state condition $\phi$ that holds after executing $\pi$ and before initiating $\pi'$. This step yields a new flow $e_{\textit{new}} = n \xrightarrow{(\pi, \phi, \pi')} n'$
We then update the flow set $E$ by comparing the newly derived flow $e_{\textit{new}}$ with each existing flow $e_{cur} = n \xrightarrow{(\pi, \phi_{cur}, \pi_{cur}')} n_{cur}'$
which shares the same source functionality $n$ and source trace $\pi$.
Based on the comparison results, the following flow-related operations are applied.

\begin{itemize}
\item \textit{New Flow Creation}.
\begin{itemize} 
    \item If no existing flow shares the same source functionality $n$ and source trace $\pi$, a new flow $n \xrightarrow{(\pi, \phi, \pi')} n'$ created.
    \item If the new condition $\phi$ is logically incomparable to an existing condition $\phi_{cur}$ and leads to a different outcome ($\pi' \neq \pi_{cur}'$), a new flow $n \xrightarrow{(\pi, \phi, \pi')} n'$ created.
    \item If $\phi \models \phi_{cur}$ (i.e., $\phi$ is stronger) and $\pi' \neq \pi_{cur}'$, a new flow $n \xrightarrow{(\pi, \phi, \pi')} n'$ created.
    \item If $\phi_{cur} \models \phi$ (i.e., $\phi_{cur}$ is stronger) and $\pi' \neq \pi_{cur}'$, a new flow $n \xrightarrow{(\pi, \phi \land \neg \phi_{cur}, \pi')} n'$ created.
\end{itemize} 

\item \textit{Existing Flow's Condition Strengthening}. If $\phi \models \phi_{cur}$  and $\pi' \neq \pi_{cur}'$, besides adding a new flow with condition $\phi$, the existing flow's condition $\phi_{cur}$ is also refined to exclude $\phi$: $e_{cur}$ is updated to $n \xrightarrow{(\pi, \phi_{cur} \land \neg \phi, \pi_{cur}')} n_{cur}'$.

\item \textit{Existing Flow's Condition Weakening}. If $\phi_{cur} \models \phi$ and $\pi' = \pi_{cur}'$ (with $n' = n_{cur}'$), the existing condition $\phi_{cur}$ is generalized to the weaker condition $\phi$: $e_{cur}$ is updated to $n \xrightarrow{(\pi, \phi, \pi_{cur}')} n_{cur}'$.

\item \textit{Existing Flow's Condition Merging}. If $\phi$ and $\phi_{cur}$ are logically incomparable and $\pi' = \pi_{cur}'$ (with $n' = n_{cur}'$), the two conditions are merged by replacing $e_{cur}$ with $n \xrightarrow{(\pi, \phi \lor \phi_{cur}, \pi')} n'$.
\end{itemize}

Overall, with the above node updating and flow updating operations, the \textit{\FDG} model can be iteratively refined during continuous test scenario generation and execution.
These modules work together to enable model refinement and systematic GUI exploration, thereby completing the \textit{inter-functional-flow-oriented GUI testing} approach.

\section{Evaluation}
To evaluate the performance of our approach, we raise several research questions as follows:
\begin{itemize}
    \item \textbf{RQ1 (Effectiveness and Efficiency)}: How effective and efficient is our approach in constructing \FDG s and detecting bugs on bug benchmarks?
    \item \textbf{RQ2 (Ablation Study)}: How does the long-short-term-view-guided testing contribute to the overall effectiveness? 
    \item \textbf{RQ3 (Practicality)}: Can our approach be successfully applied to large-scale, closed-source commercial apps to uncover new functional defects?
\end{itemize}
\subsection{Experimental Setup}

\textbf{Benchmark.}
For \textbf{RQ1} and \textbf{RQ2}, we evaluate our approach on the Themis benchmark~\cite{Themis}, which initially comprises 72 reproducible crash bugs (corresponding to 72 versions) across 20 open-source apps from F-Droid~\cite{fdroid}. After filtering out versions that failed to launch or register properly, our final dataset includes \textbf{50} usable app versions along with their corresponding crash bugs. 
For \textbf{RQ3}, we evaluate our approach on 52 popular commercial apps from two official marketplaces, Google Play~\cite{googleplay} and HUAWEI AppGallery~\cite{huawei}, spanning 13 distinct categories, including entertainment, health, social, travel, etc. We filter out apps that (1) repeatedly crash during startup or normal interaction, (2) cannot be exercised by one or more tools due to incompatibility, or (3) require mandatory registration/login that cannot be bypassed or automated with scripts.

\textbf{Baseline.} We compare our tool with five state-of-the-art GUI testing tools, i.e., Droidbot~\cite{droidbot}, Fastbot~\cite{Fastbot}, Humanoid~\cite{Humanoid}, GPTDroid~\cite{gptdroid}, and VisionDroid~\cite{visiondroid}, in which the first three tools are representative traditional GUI testing tools that respectively implement random-based, model-based, and learning-based exploration strategies. The latter two are recent LLM-based testing tools that directly leverage LLMs for exploration and testing.
To ensure a fair and advanced baseline, each baseline tool is further enhanced with a state-of-the-art, LLM-based GUI testing booster. The three traditional exploration tools are augmented with LLMDroid~\cite{llmdroid}, a framework that integrates code coverage feedback to dynamically adjust the exploration strategy, thereby improving the depth and diversity of testing. While the two native LLM-based testing tools are augmented with MemoDroid~\cite{memodroid}, a system designed to enhance LLM-driven testing through a multi-layered memory mechanism that supports the retention of execution traces, summarization of exploration experience, and strategic reuse across sessions. 

We deploy all baselines and our approach on a 64-bit Ubuntu 24.04 machine (64 cores, Intel CPU) and evaluate them using Android emulators. We run open-source apps on Google Android 7.1 emulators and commercial apps on Google Android 15 emulators; all emulators share the same configuration.
To facilitate file access from apps, we preload the SD card with various types of external files. 
Following common practice~\cite{droidbot,gptdroid,visiondroid}, for each bug that requires login, we register a dedicated account and implement the corresponding login script; before each run, we reset the account data to avoid interference. To ensure fair and reasonable resource usage, we allocate 60 minutes to each tool per app, consistent with prior GUI testing studies~\cite{droidbot,Themis,gptdroid,visiondroid}. We execute each tool three times and report the best performance to mitigate potential randomness. 
All tools (including the baselines and {\tool}) employ GPT‑4o~\cite{openai_gpt4o} as the backend MLLM, ensuring a consistent and fair comparison across methods.


\subsection{Effectiveness and Efficiency}
Table~\ref{tab:themis} demonstrates the performance of {\tool} and baseline approaches, which shows that {\tool} outperforms all baseline tools across every metric. In terms of tools' exploration ability, {\tool} achieves the highest average activity coverage (\textbf{0.55}) and code coverage (\textbf{0.40}), surpassing the best-performing baseline (VisionDroid with MemoDroid (VD+MD), 0.43 activity coverage and 0.38 code coverage) by a significant margin. This represents a relative improvement of approximately 28\% in activity coverage and 5\% in code coverage. 
This indicates that the higher coverage achieved by {\tool} expands the exploration scope of the app, thereby increasing the likelihood of detecting potential bugs.

Regarding bug detection, {\tool} demonstrates the strongest performance in finding crash bugs, uncovering \textbf{49} in total. This significantly exceeds the $24$ $(96\%)$ crash bugs found by the best baseline (VD+MD). 
Furthermore, our approach exhibits a unique advantage in detecting \textit{non-crash functional bugs}. The first three baselines (DB+LD, FB+LD, HM+LD) are built upon traditional GUI exploration tools, which lack the mechanism to detect or assert functional correctness, and thus found zero non-crash bugs. The other two baselines (GD+MD, VD+MD), augmented with advanced LLMs for semantic understanding, begin to show a nascent capability in this regard, finding 4 and 5 non-crash bugs, respectively. {\tool} significantly outperforms them, discovering \textbf{13} non-crash bugs. There are \textbf{62} bugs found by {\tool} in total.
This superior performance directly validates the effectiveness of our inter‑functional-flow-oriented testing methodology. 
\begin{table}[!h]
\setlength{\abovecaptionskip}{5pt}
\setlength{\belowcaptionskip}{-5pt}
\footnotesize
\centering
\caption{Performance of Coverage and Bug Detection.}
\label{tab:themis}
{\renewcommand{\arraystretch}{1.15}%
\rowcolors{2}{black!2}{white}%
\arrayrulecolor{black!25}%
\begin{tabularx}{0.94\linewidth}{|l|*{3}{>{\centering\arraybackslash}X|}*{2}{>{\centering\arraybackslash}X|}>{\centering\arraybackslash}X|}
\hline
\rowcolor{black!6}
\diagbox[width=9.5em,height=2.2em]{\textbf{Metric}}{\textbf{Tool}} & \textbf{DB+LD} & \textbf{FB+LD} & \textbf{HM+LD} & \textbf{GD+MD} & \textbf{VD+MD} & \textbf{FuncDroid} \\
\hline
Avg. Coverage$_{activity}$  & 0.31 & 0.36 & 0.36 & 0.42 & 0.43 & \textbf{0.55} \\
Avg. Coverage$_{code}$ & 0.25 & 0.29 & 0.23 & 0.36 & 0.38 & \textbf{0.40} \\
\hline
\#Bug$_{crash}$ & 12 & 15 & 15 & 22 & 25 & \textbf{49} \\
\#Bug$_{non-crash}$ & -- & -- & -- & 4 & 5 & \textbf{13} \\
\hline
\#Bug$_{total}$ & 12 & 15 & 15 & 26 & 30 & \textbf{62} \\
\hline
\end{tabularx}}
\vspace{0.1in}

\raggedright\small\textit{The tools: \textbf{DB+LD} is Droidbot with LLMDroid, \textbf{FB+LD} is Fastbot with LLMDroid, \textbf{HM+LD} is Humanoid with LLMDroid, \textbf{GD+MD} is GPTDroid with MemoDroid, and \textbf{VD+MD} is VisionDroid with MemoDroid.}
\end{table}

\begin{figure}[!t]
\setlength{\abovecaptionskip}{5pt}
\setlength{\belowcaptionskip}{-5pt}
    \begin{minipage}[t]{0.55\linewidth}
        \centering
        \includegraphics[width=\linewidth]{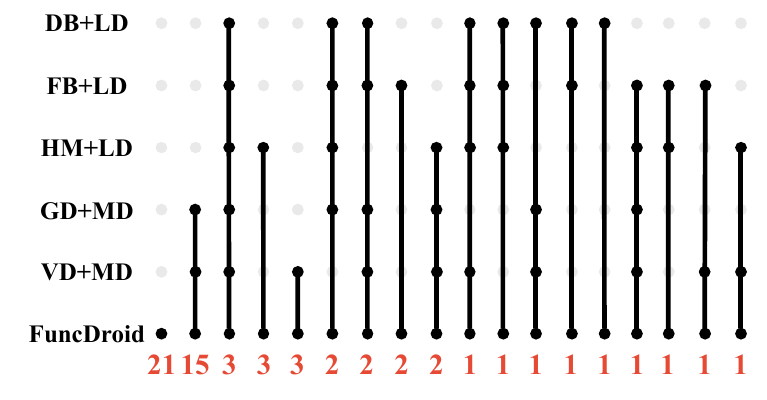}
        
         \caption{Detected Bug Overlap across Different Tool Combinations. }
        \label{fig:bug-set}
    \end{minipage}
    \hfill
    \begin{minipage}[t]{0.4\linewidth}
        \includegraphics[width=\linewidth]{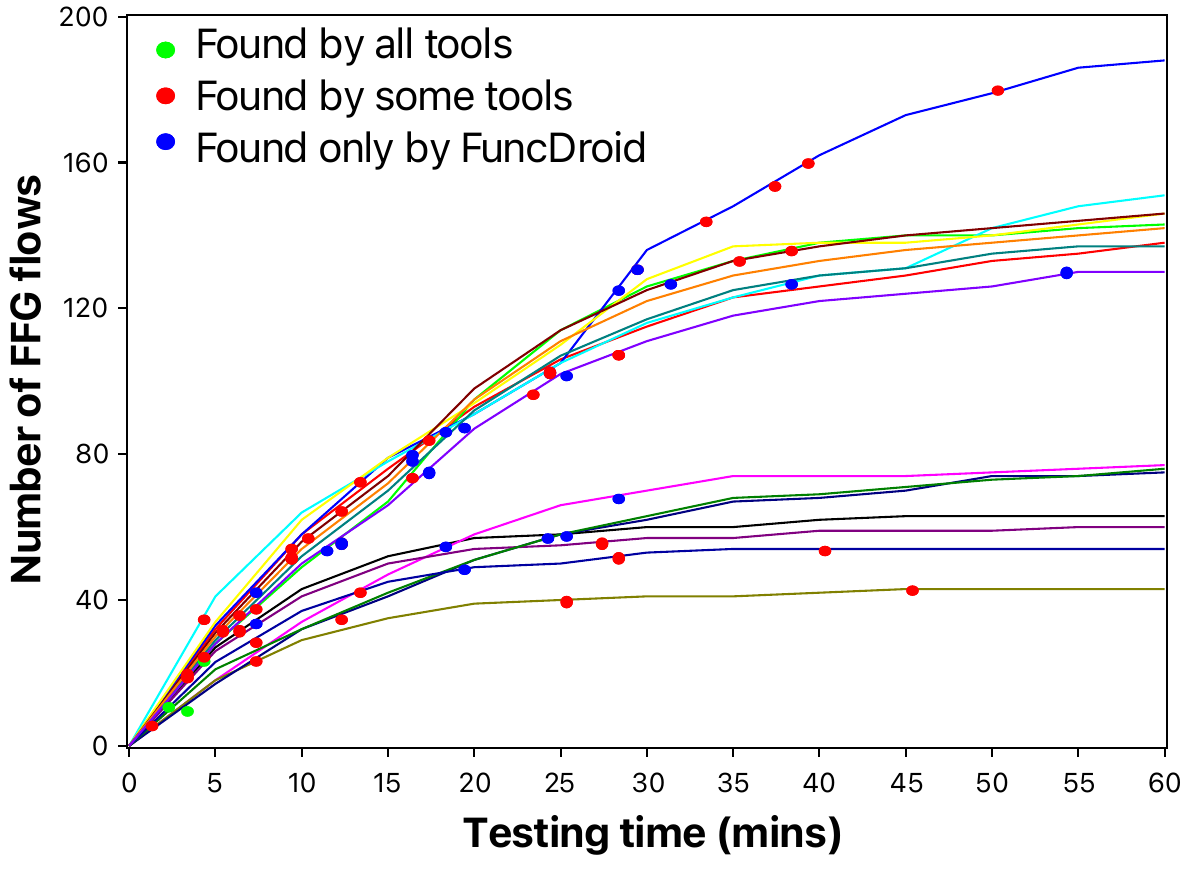}
        \caption{Number of Flows and Detected Bugs across Apps with Varying Time. }
        \label{fig:bug-flow}
    \end{minipage}
\end{figure}

Figure~\ref{fig:bug-set} summarizes the number of bugs in the overlaps detected by different tool combinations using a simplified UpSet plot.
The results indicate that only {\tool} successfully uncovered all bugs in the benchmark,
including \textbf{21 bugs} that were uniquely detected by it.
This indicates that our approach not only achieves higher quantitative results but also discovers a distinct class of defects that remain hidden to other testing techniques. Furthermore, the analysis reveals that the two LLM‑based baselines (GD+MD and VD+MD), together with {\tool}, jointly found 15 bugs that the three traditional testing tools (DB+LD, FB+LD, HM+LD) missed. Even after excluding the 4 non‑crash bugs from this set, 11 crash bugs were still only uncovered by the LLM‑based tools. This illustrates that while LLM‑enhanced traditional tools (LLMDroid) improve over purely heuristic explorers, their capability remains substantially limited compared to a truly LLM-based approach. 
We further analyze the relationship between the number of explored flows and detected bugs over time in each app, as shown in Figure~\ref{fig:bug-flow}.
The results show that {\tool} detects more bugs as the number of explored flows increases.
The blue markers on the line indicate the occurrence of unique bugs detected by {\tool}.
Notably, {\tool} is able to uncover unique bugs within the first 10 minutes,
benefiting from sufficient exploration of inter-functional flows.
As more flows are explored, {\tool} continues to identify additional unique bugs.



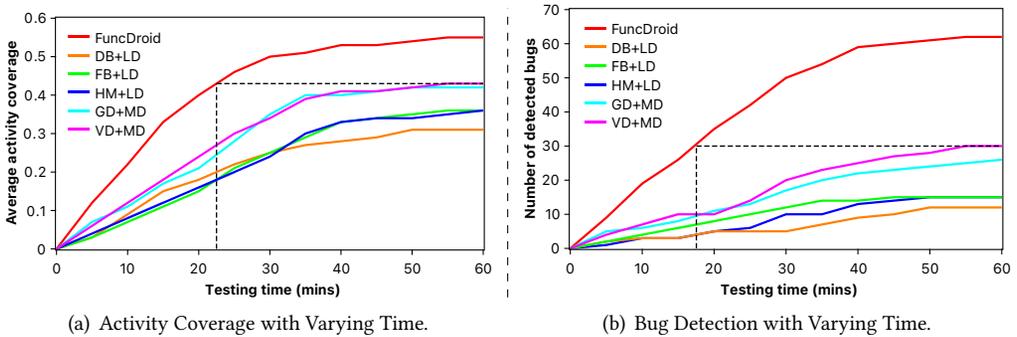
\begin{figure}[htbp]
\setlength{\abovecaptionskip}{5pt}
\setlength{\belowcaptionskip}{-5pt}
    \centering
    \subfigure[Activity Coverage with Varying Time.]{
        \label{fig:bug-example-b}
        \centering
        \includegraphics[width=0.47\linewidth]{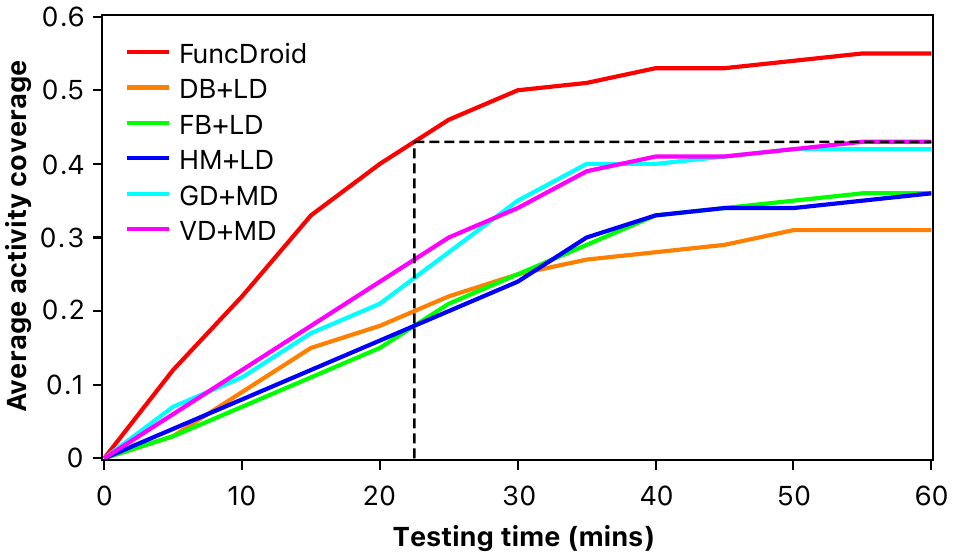}
        }
    \begin{tikzpicture}
        \draw[dashed] (0,-1.8cm) -- (0, 2cm); 
    \end{tikzpicture}
    \subfigure[Bug Detection with Varying Time.]{
        \label{fig:bug-example-a}
        \centering
        \includegraphics[width=0.47\linewidth]{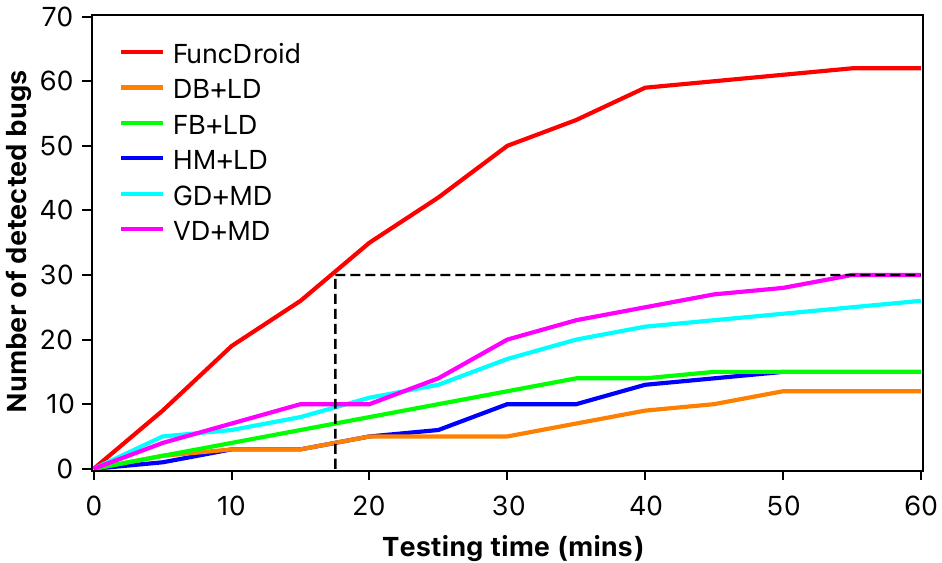}
        }
    \vspace{-0.2cm}
    \caption{Performance of Coverage and Bug Detection Ability with Varying Time. }
    \label{fig:bug-coverage}
\end{figure}

Moreover, to evaluate the efficiency of our approach, we compare {\tool} with the five baseline tools by measuring their activity coverage and the number of detected bugs over varying time. Figure~\ref{fig:bug-coverage} shows that {\tool} consistently maintains higher values in both metrics at any given time point. 
Notably, as indicated by the dashed lines, for both activity coverage and the number of detected bugs,
{\tool} reaches, at around 20 minutes, the peak performance that the best-performing baseline attains only at 60 minutes.
Additionally, we observe that the two LLM‑based baselines (GD+MD and VD+MD) also outperform the three traditional tools in this efficiency comparison. Although these LLM‑based tools incur additional overhead from interacting with the LLM, the experimental results indicate that the functionality awareness provided by the LLM yields a net positive gain in exploration speed and bug‑finding rate, allowing them to reach meaningful coverage and detect bugs faster than traditional exploration strategies.
In summary, this demonstrates that {\tool} not only achieves superior final results but also attains high effectiveness much more rapidly. This provides a strong affirmative answer to \textbf{RQ1}.

\subsection{Ablation Study}





\begin{table}[!htbp]
\setlength{\abovecaptionskip}{5pt}
\setlength{\belowcaptionskip}{-5pt}
\footnotesize
\centering
\caption{Results of the Ablation Study on {\tool}'s Long-Short-Term
View Testing}
\label{tab:ablation}
{\renewcommand{\arraystretch}{1.15}%
\rowcolors{2}{black!2}{white}%
\arrayrulecolor{black!25}%
\begin{tabularx}{0.9\linewidth}{|l|*{4}{>{\centering\arraybackslash}X|}}
\hline
\rowcolor{black!6}
\textbf{Config} & \textbf{Avg. \#Flow} & \textbf{\#Bug$_{total}$} & \textbf{\#Bug$_{crash}$} & \textbf{\#Bug$_{non-crash}$} \\
\hline
{\textbf{\tool}} & \textbf{105} & \textbf{62} & \textbf{49} & \textbf{13} \\
\hline
$w/o$ LTV-Func & 71 $(\downarrow 32\%)$ & 48 $(\downarrow 23\%)$ & 38 $(\downarrow 22\%)$ & 10 $(\downarrow 23\%)$\\
$w/o$ LTV-Flow & 89 $(\downarrow 15\%)$ & 50 $(\downarrow 19\%)$ & 39 $(\downarrow 20\%)$ & 11 $(\downarrow 15\%)$\\
$w/o$ LTV & 65 $(\downarrow 38\%)$ & 43 $(\downarrow 31\%)$ & 34 $(\downarrow 31\%)$ & 9 $(\downarrow 31\%)$\\
\hline
$w/o$ STV-Single-Flow & 97 $(\downarrow 8\%)$ & 46 $(\downarrow 26\%)$ & 37 $(\downarrow 24\%)$ & 9 $(\downarrow 31\%)$\\
$w/o$ STV-Cross-Flow & 101 $(\downarrow 4\%)$ & 49 $(\downarrow 21\%)$ & 39 $(\downarrow 20\%)$ & 10 $(\downarrow 23\%)$\\
$w/o$ STV & 94 $(\downarrow 10\%)$ & 39 $(\downarrow 37\%)$ & 33 $(\downarrow 33\%)$ & 6 $(\downarrow 54\%)$\\
\hline
$w/o$ LSTV & 58 $(\downarrow 45\%)$ & 27 $(\downarrow 56\%)$ & 23 $(\downarrow 53\%)$ & 4 $(\downarrow 69\%)$\\
\hline
\end{tabularx}}

\end{table}

In this part, we perform extra ablation studies to evaluate the contributions of the \textit{long-short-term-view-guided testing} approach. 
For a more fine-grained comparison, the long‑term view (LTV) test scenario generation is  divided into scenarios for refining functionality definitions and flow conditions, while the short‑term view (STV) test scenario generation is divided into scenarios for single-flow and cross-flow. 
The results are shown in Table~\ref{tab:ablation}, which confirm that each component independently contributes to both flow coverage and bug detection.

Disabling the \textbf{Test Scenarios Generation for Functionality Definitions ($w/o$ LTV‑Func)} phase within LTV  reduces the average number of flows in {\FDG} by 32\% and the total bugs found by 23\% (from 62 to 48). This indicates that refining the functionality definitions through completeness and independence is essential for building more accurate functionalities, which in turn yields a richer set of flows.
Similarly, disabling the \textbf{Test Scenarios Generation for Flow Conditions ($w/o$ LTV‑Flow)} phase within LTV reduces the average number of flows by 15\% and total bugs by 19\%. This result highlights the importance of sharpening state preconditions $\phi$; without such refinement, the model fails to capture precise enabling conditions, leading to both fewer flows and missed bugs. Although it contributes fewer new flows than LTV‑Func, the flows it refines become more reliable for detecting bugs. Together, these components are critical for constructing a high‑fidelity {\FDG}: one expands the flows, while the other ensures its logical accuracy, together forming a robust foundation for effective testing.
Disabling the entire \textbf{Long‑Term View Testing ($w/o$ LTV)} reduces the average number of flows by 38\% and total bugs by 31\%. This underscores that the combined effort of refining both functionalities and flows is critical for constructing a high‑fidelity {\FDG}, which serves as the foundation for effective testing.

Disabling the \textbf{Single-Flow Metamorphic Testing ($w/o$ STV-Single)} phase within STV reduces the total bugs found by 26\% (from 62 to 46), with a moderate decrease of 8\% in the average number of flows. This indicates that systematically varying how a flow’s condition is satisfied is essential for effectively uncovering crash bugs triggered by minor GUI variations. Similarly, disabling the \textbf{Cross-Flow Metamorphic Testing ($w/o$ STV-Cross)} phase reduces the total number of bugs by 21\% and the average number of flows by 4\%, highlighting the importance of testing indirect data dependencies across functionalities. Without exploring these shared data dependencies, the testing misses subtle non-crash bugs that arise from complex, multi-functional interactions. Disabling the entire \textbf{Short-Term View Testing ($w/o$ STV)} reduces the total bugs by 37\% and the average number of flows by 10\%. This underscores that the combined power of both metamorphic testing strategies is critical for intensive, in-depth exploration based on the current {\FDG} model, directly driving the detection of a wide spectrum of interaction bugs, especially non-crash bugs, which see a 54\% decline.

Disabling the combined \textbf{Long-Short Term View Testing ($w/o$ LSTV)}---which removes both the model-refinement capability of LTV and the intensive testing capability of STV---reduces the average number of flows by 45\% and the total bugs by 56\%. This configuration leaves only the initial {\FDG} bootstrap without any iterative refinement or exploitation. The drastic decline in flow count highlights the indispensable role of LTV in refining the {\FDG}, while the severe drop in bug detection, with non-crash bugs plummeting by 69\%, underscores that STV is essential for bug detection based on the {\FDG}.

\subsection{Practicality}

Finally, to assess the practicality of {\tool} on complex real-world apps, we conduct an extensive evaluation on a benchmark consisting of 52 highly downloaded commercial mobile apps spanning 13 categories.
Our tool {\tool} successfully detects \textbf{18} non-crash bugs on this benchmark.
Table~\ref{tab:google} presents detailed information about the apps in which these bugs are found.
We further conduct the same evaluation using the two LLM-based baselines from RQ1, namely GPTDroid and VisionDroid, both augmented with MemoDroid.
Neither baseline discovers any new bugs beyond those already reported.
As shown in the last two columns in Table~\ref{tab:google}, GPTDroid with MemoDroid (GD+MD) identifies only \textbf{4} bugs, while VisionDroid with MemoDroid (VD+MD) detects \textbf{6} bugs.
Notably, all bugs found by GD+MD are also covered by VD+MD.

The bug reporting for commercial apps is often challenging, but we still attempted to contact developers to validate the bugs via in-app feedback systems, email communications, etc.
Among the \textbf{18} bugs reported by {\tool}, \textbf{2} have been fixed by developers, \textbf{2} have been confirmed, and \textbf{3} were from apps without accessible bug-reporting channels; notably, none of the reported bugs were rejected. Their latest status is also shown in Table~\ref{tab:google}. The video demonstrations of these bugs found by {\tool} can be viewed online~\footnote{\url{https://www.youtube.com/playlist?list=PLB90RF1cJFTPYPATJU7U-wjkr_6torGjC}}.

\begin{table}[!htbp]
\setlength{\abovecaptionskip}{5pt}
\setlength{\belowcaptionskip}{-5pt}
\centering
\footnotesize
\caption{Bugs Detected by {\tool} on Commercial Apps.}
\label{tab:google}
{\renewcommand{\arraystretch}{1.15}%
\rowcolors{2}{black!2}{white}%
\arrayrulecolor{black!25}%
\begin{tabular}{|c|c|c|c|c|c|c|c|}
\hline
\rowcolor{black!6}
\textbf{Id} & \textbf{App Name} & \textbf{Category} & \textbf{Download} & \textbf{Size (MB)} & \textbf{Status} & \textbf{GD+MD} & \textbf{VD+MD} \\
\hline
1 & YouTuBe Kids & Entertainment & 500M+ & 27 & waiting & \xmark & \xmark \\
2 & PiPiXia & Entertainment & 100M+ & 107 & fixed & \xmark & \xmark \\
3 & ZuiYou & Entertainment & 100M+ & 159 & fixed & \xmark & \xmark \\
4 & Blood Pressure & Health & 10M+ & 26 &waiting  & \xmark & \xmark \\
5 & MeiYou & Health & 100M+ & 175 & unreportable & \cmark & \cmark \\
6 & AnJuKe & Lifestyle & 400M+ & 88 & waiting & \xmark & \xmark \\
7 & Live Earth Map & Navigation & 1M+ & 28 &waiting  & \xmark & \xmark \\
8 & KugouMusic & Music & 1B+ & 162 & waiting & \xmark & \xmark \\
9 & FOX LOCAL & News & 500K+ & 18 &waiting  & \xmark & \xmark \\
10 & Substack & News & 5M+ & 31 & waiting & \cmark & \cmark \\
11 & PDF File Reader & Productivity & 5M+ & 24 &waiting  & \xmark & \xmark \\
12 & Instagram & Social & 5B+ & 97 &waiting  & \cmark & \cmark \\
13 & Wechat & Social & 1B+ & 243 & confirmed & \xmark & \xmark \\
14 & Victory+ & Sports & 100K+ & 39 & confirmed & \xmark & \cmark \\
15 & QR Code\&Barcode Scanner & Tools & 1M+ & 20 & unreportable & \xmark & \xmark \\
16 & Greyhound: Buy Bus Tickets & Travel & 1M+ & 14 &waiting  & \xmark & \cmark \\
17 & Allegiant & Travel & 5M+ & 54 &waiting  & \cmark & \cmark \\
18 & Qunar Travel & Travel & 200M+ & 82 & unreportable & \xmark & \xmark \\
\hline
\end{tabular}}

\end{table}
\section{Related Work}
Automated GUI testing for mobile applications has evolved through several distinct phases, broadly categorized by their primary objective: maximizing exploration coverage or understanding and testing application functionality.
Accordingly, we discuss related work from these two perspectives.

\textbf{Coverage-oriented GUI Testing.} 
The primary goal of this line of work is to explore as many distinct states and activities within an application as possible.
Early random testing methods~\cite{monkey, droidbot,time-machine,random1}  and fuzzing methods~\cite{fuzzing1,fuzzing2,fuzzing3} generate unpredictable event sequences.
While efficient and simple, their lack of structural or semantic guidance often results in low coverage and redundant exploration.
To mitigate these limitations, model-based testing techniques~\cite{droidbot,APE,Sapienz,TrimDroid,Stoat,ComboDroid,WTG,grey-box,Fastbot,hacmony,afmm,atg} construct explicit representations of the GUI, such as state transition graphs, to guide systematic exploration.
These models help avoid redundant states but often suffer from state explosion and, more critically, remain agnostic to the semantic purpose of the interfaces they traverse.
To further enhance exploration intelligence, learning-based testing approaches~\cite{Humanoid,curiosity-driven,Q-network,Reinforcement} leverage techniques from reinforcement learning and deep learning.
By learning interaction patterns from data, these methods prioritize promising actions and generate test cases that align more closely with actual user behaviors. However, their effectiveness is constrained by a heavy reliance on the quality and scale of training datasets, which limits generalization and makes them difficult to adapt to rapidly evolving application functionalities.

\textbf{Functionality-oriented GUI Testing.} 
This line of work shifts focus from covering states to understanding and executing user-centric tasks. 
To effectively test functionality, property‑based testing method ~\cite{Genie,Odin,kea} systematically generates input sequences to verify specific semantic properties of the apps. However, they typically lack a deeper understanding of functional behaviors.
The advent of LLMs (especially MLLMs) has been pivotal in bridging this gap, enabling a leap in semantic interpretation of GUI content~\cite{gptdroid,llmdroid,visiondroid,AppAgent,droidbot-gpt,VisionTasker,autodroid,MobileAgent,MobileExperts,memodroid}. Early works such as AppAgent~\cite{AppAgent} and AutoDroid~\cite{autodroid} focus on multimodal perception and task planning. 
However, these methods still focus on the interface or widget level and do not model the functional semantics or logic between functions. GPTDroid~\cite{gptdroid} and VisionDroid~\cite{visiondroid}, employ LLMs to directly interpret GUI pages, summarize page sequences into reusable functional patterns, and then apply these patterns to guide further exploration. Although such approaches improve semantic comprehension and planning, they still focus primarily on individual functionalities. Frameworks such as LLMDroid~\cite{llmdroid} and MemoDroid~\cite{memodroid} enhance existing testing tools (whether traditional or LLM‑based) by constructing knowledge bases that capture app‑specific functional characteristics. These frameworks distill common interaction patterns and effective exploration strategies from historical testing sessions to steer exploration more efficiently.

In summary, while coverage-oriented methods provide broad but shallow exploration, and modern LLM-based methods enable functionality-aware task execution, the field still lacks a model that explicitly represents and facilitates the systematic testing of inter-functional flows, which is the precise gap our work addresses with the inter-functional flow oriented GUI testing method.

\section{Conclusion}


Mobile application testing plays a critical role in ensuring the reliability and quality of software systems that are deeply embedded in everyday life.
However, the key limitation of functionality-oriented GUI testing lies in the absence of a systematic modeling and testing approach for inter-functional interactions.
This work addresses this limitation by providing a scalable and practical solution that bridges the capability gap in detecting complex, interaction-dependent bugs.
By focusing on the coordinated interactions among functionalities at a higher level of abstraction, this work advances automated GUI testing beyond traditional coverage- or single-functionality-centric paradigms.
Specifically, we propose a novel \textbf{Functional Flow Graph} (\FDG) to explicitly model inter-functional interactions, and introduce an \textit{inter-functional-flow-oriented testing} methodology that iteratively refines the {\FDG} while guiding testing toward deep interaction paths.
Extensive experimental results demonstrate the effectiveness, efficiency, and practical applicability of our approach in detecting both crash and non-crash bugs.
Looking ahead, future work will focus on extending this approach to additional mobile platforms (e.g., iOS, HarmonyOS) and enhancing the knowledge base for non-crash bug detection, further improving the overall bug detection capability.
\section*{Data Availability}
The implementation of {\tool} and all experimental datasets and results are publicly available at \url{https://github.com/daxiami233/FuncDroid}. The bugs discovered by {\tool} are showcased at \url{https://fim-bugs.sites.veilaxis.com}.

\bibliographystyle{ACM-Reference-Format}
\bibliography{ref}


\end{document}